\documentclass{ifacconf}

\usepackage{graphicx}      % include this line if your document contains figures
\usepackage{natbib}        % required for bibliography
\usepackage{balance}
\usepackage{xspace}
\usepackage{bbm}
\usepackage{rotating}
\usepackage{url}
\usepackage{csquotes}
\usepackage{subfig}
\usepackage{paralist}
\usepackage{multirow}
\usepackage{multicol}
\usepackage{amsmath,amssymb,mathtools}
\usepackage{booktabs}
\usepackage{xfrac}
\usepackage{array}
\usepackage{algorithm}
\usepackage[noend]{algpseudocode}
\usepackage{textcomp}
\usepackage{siunitx}
\usepackage{microtype}
\usepackage[usenames,dvipsnames]{xcolor}
\usepackage{pgfplots}
\pgfplotsset{compat=newest,unit code/.code={\si{#1}},plot coordinates/math parser=false,grid style={lightgray}, ylabel right/.style={
        after end axis/.append code={
            \node [rotate=90, anchor=north] at (rel axis cs:1,0.5) {#1};
        }   
    }}
\usepgfplotslibrary{units,external,groupplots,fillbetween}
\usetikzlibrary{positioning,angles,quotes,patterns,shapes,spy, shapes.misc,backgrounds}
\tikzstyle{block} = [draw, rectangle, minimum height=2em, minimum width=5em]
\tikzstyle{addon} = [draw, rectangle, rounded corners]
\tikzstyle{pinstyle} = [pin edge={<-,thin,black}]
\tikzstyle{pinstyle2} = [pin edge={->,thin,black}]
\tikzstyle{mult} = [draw, isosceles triangle]
\tikzstyle{circ} = [draw, circle]
\tikzstyle{coord} = [coordinate]
\tikzstyle{circ2} = [draw, circle,minimum width=3pt, inner sep=0]
\tikzset{>=latex}
\tikzset{radiation/.style={{decorate,decoration={expanding
waves,angle=90,segment length=4pt}}}}
\usetikzlibrary{math}
\tikzmath
{
  function symlog(\x,\a){
    \yLarge = ((\x>\a) - (\x<-\a)) * (ln(max(abs(\x/\a),1)) + 1);
    \ySmall = (\x >= -\a) * (\x <= \a) * \x / \a ;
    return \yLarge + \ySmall ;
  };
  function symexp(\y,\a){
    \xLarge = ((\y>1) - (\y<-1)) * \a * exp(abs(\y) - 1) ;
    \xSmall = (\y>=-1) * (\y<=1) * \a * \y ;
    return \xLarge + \xSmall ;
  };
}

\graphicspath{{./figures/}}

\usepackage{scalerel,stackengine}

\makeatletter
\newcommand{\myitem}[1]{%
\item[#1]\protected@edef\@currentlabel{#1}%
}
\makeatother

\DeclareSIUnit{\belmilliwatt}{Bm}
\DeclareSIUnit{\dBm}{\deci\belmilliwatt}

\DeclareMathOperator*{\E}{\mathbb{E}}

\DeclareMathOperator*{\Tr}{Tr}
\newcommand{\norm}[1]{\lVert#1\rVert}

\newcommand{\x}[1]{\tilde{x}_{#1}}

\let\originalleft\left
\let\originalright\right
\renewcommand{\left}{\mathopen{}\mathclose\bgroup\originalleft}
\renewcommand{\right}{\aftergroup\egroup\originalright}

\newcommand\figref[1]{Fig.~\ref{#1}}

\newcommand{\eg}{e.g.,\xspace}
\newcommand{\ie}{i.e.,\xspace}

\newcommand{\cf}{cf.\@\xspace}
\newcommand{\capt}[1]{\mdseries{\emph{#1}}}

\usepackage{ifthen}
\newboolean{authnotes}

\setboolean{authnotes}{true}
\ifthenelse{\boolean{authnotes}}
{
\newcommand{\am}[1]{\footnote{{\bf\color{blue!70!black} Alon: #1}}}
\newcommand{\db}[1]{\footnote{{\bf\color{green!50!black} Dominik: #1}}}
\newcommand{\st}[1]{\footnote{{\bf\color{purple!90!black} Sebastian: #1}}}
\newcommand{\mt}[1]{\footnote{{\bf\color{orange!50!black} Matteo: #1}}}
}
{
\newcommand{\am}[1]{}
\newcommand{\db}[1]{}
\newcommand{\st}[1]{}
\newcommand{\mt}[1]{}
}

\usepackage{fancyhdr}
\newcommand{\mytitle}{\textbf{Accepted final version.}
To appear in \textit{Proc. of the IFAC World Congress, 2023}.\\
\copyright 2023 the authors. This work has been accepted to IFAC for publication under a Creative Commons Licence CC-BY-NC-ND}
\begin{document}
\begin{frontmatter}

\title{On the trade-off between event-based\\ and periodic state estimation\\ under bandwidth constraints} 

\thanks[footnoteinfo]{This research was  financially supported by the project \emph{NewLEADS - New Directions in Learning Dynamical Systems} (contract number: 621-2016-06079), funded by the Swedish Research Council. The research was also supported by \emph{Kjell och Märta Beijer Foundation}.}

\author[First,Second]{Dominik Baumann} 
\author[Second]{Thomas B.\ Sch\"{o}n} 
% \author[Third]{Third C. Author}

\address[First]{Department of Electrical Engineering and Automation, Aalto University, Espoo, Finland (e-mail: dominik.baumann@aalto.fi)}
\address[Second]{Department of Information Technology, Uppsala University, Uppsala, Sweden (e-mail: thomas.schon@ it.uu.se).}
% \address[Second]{Department of Electrical Engineering and Automation, Aalto University, Espoo, Finland (e-mail: dominik.baumann@aalto.fi)}
% \address[Third]{Electrical Engineering Department, 
%    Seoul National University, Seoul, Korea, (e-mail: author@snu.ac.kr)}

\begin{abstract}                % Abstract of not more than 250 words.
Event-based methods carefully select when to transmit information to enable high-performance control and estimation over resource-constrained communication networks.
However, they come at a cost. 
For instance, event-based communication induces a higher computational load and increases the complexity of the scheduling problem.
Thus, in some cases, allocating available slots to agents periodically in circular order may be superior.
In this article, we discuss, for a specific example, when the additional complexity of event-based methods is beneficial.
We evaluate our analysis in a synthetical example and on 20 simulated cart-pole systems.
\end{abstract}


\end{frontmatter}

%%%%%%%%%%%%%%%%%%%%%%%%%%%%%%%%%%%%%%%%%%%%%
% Added after final submission
%%%%%%%%%%%%%%%%%%%%%%%%%%%%%%%%%%%%%%%%%%%%%
\thispagestyle{fancy}   % final submitted: empty
\pagestyle{empty}
%===============================================================================

%% There are a number of predefined theorem-like environments in
%% ifacconf.cls:
%%
%% \begin{thm} ... \end{thm}            % Theorem
%% \begin{lem} ... \end{lem}            % Lemma
%% \begin{claim} ... \end{claim}        % Claim
%% \begin{conj} ... \end{conj}          % Conjecture
%% \begin{cor} ... \end{cor}            % Corollary
%% \begin{fact} ... \end{fact}          % Fact
%% \begin{hypo} ... \end{hypo}          % Hypothesis
%% \begin{prop} ... \end{prop}          % Proposition
%% \begin{crit} ... \end{crit}          % Criterion

\section{Introduction}

Remote estimation and control over communication networks require us to account for the constraints that communication networks impose.
For instance, exceeding the bandwidth by transmitting information at high periodic rates can cause increasing delays and a higher probability of losing messages \citep{hespanha2007survey}.
In response, event-based methods \citep{heemels2012introduction,grune14event,miskowicz15event,lemmon11networked} break the periodic paradigm and let systems transmit information ``only when necessary.''
Whether or not information exchange is necessary is usually decided by a so-called \emph{trigger} based on, for instance, the estimation error.
While such methods can significantly reduce the number of transmitted messages \citep{trimpe2011experimental,araujo2013system,dolk2017event}, they require additional computations, and communication slots need to be allocated to systems online.
Thus, in some cases, it may be beneficial to simply let systems communicate by turns instead.
In this article, we examine this trade-off for a recently proposed event-based method and close with a discussion on further research in this direction.

The area of event-based estimation and control has seen substantial growth since the beginning of the century.
Nevertheless, studies that actually integrate existing methods into wireless networks are still relatively sparse.
In particular, bandwidth savings are only shown in very few cases \citep{baumann2019control}.
Two main challenges are responsible.
Most event-based methods make instantaneous decisions about whether or not to transmit a message.
Thus, the communication system has no time to reallocate communication slots and, essentially, needs to enable communication for all systems at all time slots.
The systems can then spontaneously decide whether or not to use the slot.
Self-triggered control and estimation methods \citep{velasco2003self,wang2009self,mazo2010iss} account for that by announcing the next time a system needs to communicate at the current communication instant.
This leaves the communication system time to reallocate resources dynamically and can enable resource savings \citep{araujo2013system,ma2018efficient,baumann2019control}, overcoming the first challenge.
The second challenge lies in the binary nature of both event-based and self-triggered methods.
Systems decide to either communicate or not.
Thus, in the worst case, all systems may simultaneously announce a need for communication, making it again necessary that all systems can transmit information at once.
In wireless networks, this may be impossible and cause the network to break down.
A remedy to this problem is contention resolution algorithms \citep{ramesh2016performance,balaghi2018decentralized,mamduhi2017error,molin2011optimal} that, for instance, award communication slots to the systems with the highest error.
In a similar direction, predictive triggering \citep{mastrangelo2019predictive,mager2021scaling} aims at predicting communication priorities and scheduling communication according to those priorities.
To the best of our knowledge, predictive triggering is the only event-based method that addresses both challenges \emph{and} has been evaluated in experiments including a real wireless network.
Thus, we will investigate this approach in the following.

In the experimental evaluation by \cite{mager2021scaling}, the authors compare predictive triggering to round-robin communication.
In round-robin communication, a periodic schedule is derived a priori letting systems communicate in turns.
While this scheme ignores the actual communication needs of each system at communication times, it does not require any online calculations.
Further, predictive triggering requires transmitting the communication priorities to derive a schedule.
This puts an additional load on the wireless network.
Thus, while round-robin communication allocates slots independent of communication needs, it can use the entire bandwidth for transmitting control messages.
In \cite{mager2021scaling}, this results in a higher number of control messages that can be transmitted with round-robin communication per round as compared to predictive triggering.

This suggests an interesting fundamental trade-off.
While predictive triggering allocates slots to the systems that are in the most desperate need of communication, it can award only fewer slots.
In this article, we examine this trade-off and derive a criterion for when predictive triggering is superior to round-robin communication and vice versa.

\section{Problem setting}

\begin{figure}
\centering
% !TEX root = ../root.tex
\tikzsetnextfilename{problem_setting}
\begin{tikzpicture}
\node[align=center, draw, text=black, cloud, cloud puffs=17, cloud puff arc=140, aspect=2.5, fill=white, minimum width=15em, minimum height=4em](nw){Wireless network};
\node[draw, rectangle, above=2.5em of nw](scheduler){Scheduler};
\node[draw, circle, above=2.5em of scheduler](minus){};
\node[draw, rectangle, left=3em of minus](sys){System $i$};
\node[draw, rectangle, right=3em of minus, align=left](loc_est){Local\\ estimator $i$};
\node[draw, rectangle, below=3em of nw, align=left](rem_est){Remote\\ estimator $i$};
\node[rectangle, minimum width=1em, minimum height=1em]at(sys|-scheduler)(switch){};

\draw[->](sys) --node[midway, above]{$x_i(k)$} (minus);
\draw[->](loc_est) --node[midway, above]{$\hat{x}_i(k)$}node[pos=0.9, below]{$-$}(minus);
\draw[->](minus) --node[midway, right]{$\hat{e}_i(k)$}(scheduler);
\draw[->](scheduler) -- (switch.east);
\draw[->](nw) -- node[midway, right]{$\hat{e}_1(k),\ldots,\hat{e}_\mathrm{N}(k)$}(scheduler);
\draw[->, dashed](nw) -- node[midway, right]{$x_i(k-1)$}(rem_est);
\draw(sys) -- node[midway, right]{$x_i(k)$} (switch.north);
\draw[->](switch.south) -- (nw.north-|switch);
\draw (switch.north) -- +(1em, -1em);

\end{tikzpicture}
\caption{The considered problem setting. \capt{We consider \(N\) systems where a remote estimator shall estimate the state of each system. Sensor information from system to estimator must be transmitted via a bandwidth-constrained network.}}
\label{fig:problem_setting}
\end{figure}
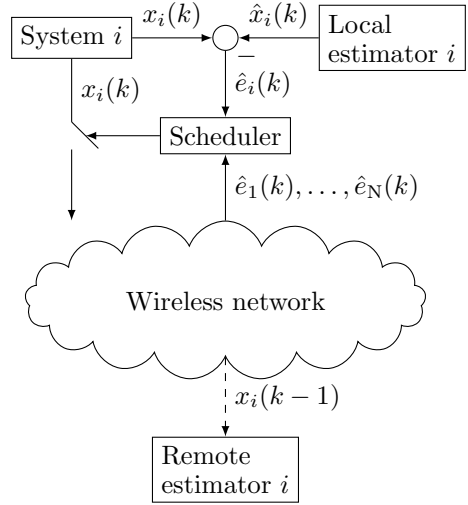

We consider linear, time-invariant dynamical systems
\begin{equation}
\label{eqn:sys}
x_i(k+1) = A_ix_i(k) + v_i(k),
\end{equation}
with discrete time index \(k\in\mathbb{N}\), system index \(i\in\{1,\ldots,N\}\), state \(x_i(k)\in\mathbb{R}^n\), process noise \(v_i(k)\sim\mathcal{N}(0,\Sigma_i)\) with variance \(\Sigma_i\), and \(A_i\) the state transition matrix of appropriate dimension.
Our goal is to estimate the state \(x_i(k)\) of each system with a remote state estimator (\cf \figref{fig:problem_setting}) whose updates are given by
\begin{equation}
\label{eqn:state_est}
\hat{x}_i(k) = \begin{cases}
A_ix_i(k-1)\quad &\text{ if }\gamma_i(k-1) = 1,\\
A_i\hat{x}_i(k-1)&\text{ if }\gamma_i(k-1) = 0.
\end{cases}
\end{equation}
We assume communication occurs in dedicated rounds with intervals of one discrete time step.
Thus, if system \(i\) is allocated a communication slot \emph{and} decides to communicate its state (\(\gamma_i(k-1)=1\)), the information will arrive one time step later at the remote estimator.
Therefore, the estimator predicts the received state one time step into the future.
In case of no communication (\(\gamma_i(k-1)=0\)), it propagates its last estimate.

The logic that makes the communication decisions is located at the sensor.
Thus, we can calculate~\eqref{eqn:state_est} not only at the estimator but also at the sensor.
Then, the sensor can compute the error \(e_i(k) = x_i(k)-\hat{x}_i(k)\), which is used in our triggering scheme.

We make the following assumption about system~\eqref{eqn:sys}.
\begin{assum}
\label{ass:noise}
For any \(e_i(k)\), we have that
\[
\norm{A_i}^2\norm{e_i(k)} + \Tr(\Sigma_i) > \norm{e_i(k)}.
\]
\end{assum}
As we will see in the error analysis, this assumption excludes systems for which the estimation error would decay over time even without communication, making communication essentially unnecessary.

We consider a setting where bandwidth is limited, \ie only \(K<N\) systems can simultaneously send their state measurements over the network.
We adopt two schemes for deciding which systems to allocate slots to: predictive triggering and round-robin communication.
For predictive triggering, we use the quadratic norm of the errors, \ie \(\norm{e_i(k)}^2\), as priority measures.
The systems exchange those priorities at the beginning of each communication round to enable informed scheduling decisions.
However, exchanging priorities also uses up some of the bandwidth, further reducing the number of communication slots.
That is, for predictive triggering, only \(K_\mathrm{pred}<K<N\) can transmit information in each time step.
We then allocate slots to the \(K_\mathrm{pred}\) agents whose quadratic error norms \(\norm{e_i(k)}^2\) are the largest.
As an alternative, we consider round-robin communication.
For round-robin communication, in the first round, agents \(\{1,\ldots,K_\mathrm{per}\}\) will receive a slot, in the second round, agents \(\{K_\mathrm{per}+1,\ldots,2K_\mathrm{per}\}\) and so on.
Since this entails no further overhead, we can use \(K_\mathrm{per}=K\) slots to transmit state information.

While predictive triggering allocates slots to the agents with the highest error, in some rounds, this might still include agents with a small error.
Similarly, during round-robin communication, systems that are actually in no need of communication may be awarded a slot.
Thus, before transmitting a message, we let each system check whether its estimation error is above a predefined threshold \(\delta\).
For systems that are awarded a slot at time \(k\), the communication decision is then defined by
\begin{equation}
\label{eqn:comm_dec}
\gamma_i(k) = \begin{cases}
1 \quad &\text{ if } \norm{e_i(k)}^2 \ge \delta,\\
0 &\text{ if } \norm{e_i(k)}^2 < \delta.
\end{cases}
\end{equation}
We here consider equal thresholds \(\delta\) among all agents, but those could also be individual.
While the round-robin and predictive triggering scheduling schemes ensure that bandwidth constraints are respected, the triggering mechanism in~\eqref{eqn:comm_dec} additionally allows for energy savings.
Such a strategy has been implemented in a real system by \cite{mager2021scaling}, and its potential for energy savings has been shown.

This article aims to find a simple criterion for when to switch between event-based and periodic communication.
This is especially interesting in settings where the system dynamics may change over time.
Recent works have, particularly in the area of event-based communication, investigated when system dynamics have changed significantly.
Then, they propose to re-identify or re-learn the dynamics \citep{solowjow2020event,baumann2019event}.
Especially then, it may be interesting to switch the communication paradigm, and we will discuss when this should be done in the following.

\section{Error analysis}
\label{sec:err_analysis}

We now analyze the estimation error in both periodic and predictive triggering and, from that, derive a decision for which of the two schemes to use based on the system dynamics, number of systems, and bandwidth constraints.

First, we derive an expression for the error dynamics.
Assume \(\gamma_i(k)=0\).
Then
\begin{align}
\label{eqn:err_dyn}
\begin{split}
e_i(k+1) &= x_i(k+1)-\hat{x}_i(k+1)\\
&= A_ix_i(k)+v_i(k)-A_i\hat{x}_i(k)\\
&= A_ie_i(k)+v_i(k).
\end{split}
\end{align} 

\subsection{Periodic control}
\label{sec:err_per}

In the case of periodic control, an error estimate is relatively easy to obtain.
Each agent will get a slot at the latest every \(T_\mathrm{per}=\text{ceil}(N/K_\mathrm{per})\) steps, where ceil is a function rounding its argument to the next higher integer.
Suppose system \(i\) was given a communication slot at time step \(k\).
If it takes the slot, the current state will be communicated, resetting the error.
Otherwise, the error was smaller than \(\delta\).
We can then get an upper bound for the expected quadratic error.
\begin{prop}
\label{prop:per}
Given the system~\eqref{eqn:sys}, estimator~\eqref{eqn:state_est}, round-robin communication, and a communication slot for system~\(i\) at time \(k\). 
Assume that the estimators are initialized with \(\norm{e_i(0)}^2<\delta\) for all \(i\).
Then, we have
\begin{align}
\label{eqn:per_quadr_err_bound}
\begin{split}
&\E[\norm{e_i(k+T_\mathrm{per})}^2] \le \\
&\norm{A_i^{T_\mathrm{per}}}^2\delta + \sum_{j=0}^{T_\mathrm{per}-1} \norm{A_i^{T_\mathrm{per}-j-1}}\Tr(\Sigma_i).
\end{split}
\end{align}
\end{prop}
\begin{pf}
Leveraging that we assume the noise to have zero mean and the Cauchy-Schwarz inequality, we have
\begin{align*}
&\E[\norm{e_i(k+T_\mathrm{per})}^2\mid e_i(k)]\\
&= \E[\norm{A_i^{T_\mathrm{per}}e_i(k) + \sum_{j=0}^{T_\mathrm{per}-1}A_i^{T_\mathrm{per}-j-1}v(j)}^2\mid e_i(k)]\\
&= \E[\norm{A_i^{T_\mathrm{per}}e_i(k)\mid e_i(k)}^2] + \E[\norm{\sum_{j=0}^{T_\mathrm{per}-1}A_i^{T_\mathrm{per}-j-1}v(j)}^2]\\
&\le \norm{A_i^{T_\mathrm{per}}}^2\norm{e_i(k)}^2 + \sum_{j=0}^{T_\mathrm{per}-1}\norm{A_i^{T_\mathrm{per}-j-1}}^2\Tr(\Sigma_i).
\end{align*}
If system \(i\) communicated at time \(k\) we reset \(e_i(k)\).
Otherwise, its squared norm is upper bounded by \(\delta\), which proves the claim.\qed
\end{pf}

\subsection{Predictive triggering}
\label{sec:err_pred}

For predictive triggering, the error analysis is slightly more involved since there are no fixed rounds at which any system~\(i\) should be awarded a slot.
However, we can intuitively derive an upper bound by assuming that the system whose error grows fastest will be the last that is assigned a slot.
To formalize this, we first define a gain.
\begin{defn}
\label{def:gain}
We define the gain of a system~\eqref{eqn:sys} as
\[
   \zeta_i = \norm{A_i}^2\delta + \Tr(\Sigma_i).
\]
\end{defn}
Then, we can upper bound the error with predictive triggering.
\begin{lem}
\label{lem:err_pred}
Given the system~\eqref{eqn:sys}, estimator~\eqref{eqn:state_est}, and an allocation mechanism based on predictive triggering. 
Assume that the estimators are initialized with \(\norm{e_i(0)}^2<\delta\) for all \(i\).
Define \(T_\mathrm{pred}\coloneqq \text{ceil}(N/K_\mathrm{pred})\).
Assume that the systems are ordered such that \(\zeta_{i-1}\le \zeta_i\).
Then, we have
\begin{align}
\label{eqn:bound_pred_trig}
\begin{split}
\E[\norm{e(T_\mathrm{pred})}^2] &\le \prod_{j=1}^{T_\mathrm{pred}}\norm{A_{(j-1)K_\mathrm{pred}+1}}^2\delta \\
&+ \sum_{j=1}^{T_\mathrm{pred}}\prod_{s=j}^{T_\mathrm{pred}}\norm{A_{(s-1)K_\mathrm{pred}+1}}^2\Tr(\Sigma_s).
\end{split}
\end{align}
\end{lem}
\begin{pf}
The intuition behind~\eqref{eqn:bound_pred_trig} is that we assume that in the first round, the \(K_\mathrm{pred}\) agents with the lowest \(\zeta_i\) communicate.
Then, the ones with the next highest.
At every time step, the error can then be upper bounded by the minimum error of the currently communicating agents since all other errors must be lower.
We have three underlying assumptions for this upper bound: \emph{(i)} ordering the communication of agents with increasing \(\zeta_i\) leads to the highest error, \emph{(ii)} no agent communicates twice, and \emph{(iii)} any error that we see after \(k=T_\mathrm{pred}\) will also be upper bounded by the upper bound for \(e(T_\mathrm{pred})\).
We now show that these assumptions are satisfied.

For the first one, assume that the agents with the highest gain communicate in the first round.
Then, we have that their error bound is
\begin{equation*}
\E[\norm{e_i(1)}^2] \le \norm{A_i}^2\delta + \Tr(\Sigma_i).
\end{equation*}
By Assumption~\ref{ass:noise}, this is upper bounded by~\eqref{eqn:bound_pred_trig}.
For all other agents \(j\), we have
\begin{align*}
\begin{split}
\E[\norm{e_j(1)}^2] &\le \norm{A_j}^2\delta + \Tr(\Sigma_j)\\
&\le \norm{A_i}^2\delta + \Tr(\Sigma_i),
\end{split}
\end{align*}
\ie their own gains provide a sharper upper bound than the gain of the systems that did communicate.
Thus, their error is also upper bounded by~\eqref{eqn:bound_pred_trig}.
At \(T_\mathrm{pred}\), assuming still that no system communicates twice, suppose that now the systems with the lowest gain communicate. 
The estimation error of those systems is upper bounded by
\begin{equation*}
\E[\norm{e_j(T_\mathrm{pred})}^2] \le \norm{A_j^{T_\mathrm{pred}}}^2\delta + \sum_{\ell=1}^{T_\mathrm{pred}}\norm{A_j^{T_\mathrm{pred}-\ell-1}}^2\Tr(\Sigma_j),
\end{equation*}
which is clearly upper bounded by~\eqref{eqn:bound_pred_trig} since the gain of this system is the lowest gain used in~\eqref{eqn:bound_pred_trig}.
In case a system with a higher gain is communicating at \(T_\mathrm{pred}\), the error gain of this system must, at some \(k\), have been upper bounded by the gain of the systems with the lowest gains (precisely when those systems communicated).
Thus, also this error must be upper bounded by~\eqref{eqn:bound_pred_trig}.

Second, assume that at some \(k\), system \(j\) receives a second slot.
Then, we have
\begin{equation*}
\E[\norm{e_j(k)}^2] \le \norm{A_j^{k-1}}^2\delta + \sum_{\ell=1}^{k-2}\norm{A_j^{k-\ell-2}}^2\Tr(\Sigma_j).
\end{equation*}
This is upper bounded by~\eqref{eqn:bound_pred_trig} since all gains of systems communicating after the first time system \(j\) communicated are higher or equal to that of \(j\) by Assumption~\emph{(i)}.
Thus, a system communicating twice would lead to a lower or equal error bound.

Consider now the last assumption.
For the first round, each agent starts with an error below \(\delta\) and we investigated the worst-case growth of that error for the worst ordering of communication.
Every system that communicated before \(T_\mathrm{pred}\) will, from then on, grow the error with its own gain since this upper bound is sharper than that of the currently communicating system, whose gain is larger.
The worst case is thus when all systems have the same gain. 
However, even then, the system communicating at \(k=1\) will either get or not get a slot at \(k=T_\mathrm{pred}+1\). 
In the first case, this would again be a period of \(T_\mathrm{pred}\), \ie the same upper bound~\eqref{eqn:bound_pred_trig}. 
In the second case, its error will be upper bounded by another agent getting an earlier slot, \ie its error will be below~\eqref{eqn:bound_pred_trig}.\qed
\end{pf}

\subsection{Trade-off}
\label{sec:trade-off}

Given the error upper bounds, we can now trade off periodic and predictive triggering.
Generally, we see that if we have a homogeneous system with \(A_i=A_j\), \(\Sigma_i=\Sigma_j\) for all \(i,j\) and \(T_\mathrm{pred}=T_\mathrm{per}\), we end up with the same upper bounds on the error.
This also implies that, since, in general, \(T_\mathrm{pred}>T_\mathrm{per}\), predictive triggering is most effective in settings with heterogeneous agents.
In periodic triggering, the maximum error is defined by the error the system with the highest gain \(\zeta_i\) accumulates over \(T_\mathrm{per}\).
In predictive triggering, this upper bound is lower if the other agents that might be assigned a slot before system \(i\) have a significantly smaller gain.

This also holds in settings with homogeneous systems and disturbances.
Suppose, for instance, a load disturbance acting on a subset of the agents, which we could model as some of the $v_i$ having a non-zero mean.
Also here, the error in periodic triggering is determined by the agent with the largest gain, while predictive triggering can assign more slots to agents that experience a load disturbance.

This resonates with the experimental results of \cite{mager2021scaling}.
In those experiments, we have 20 cart-pole systems trying to synchronize their cart positions.
The resulting multi-agent system is heterogeneous since six cart-poles are real systems from two different manufacturers, and 14 are linear simulations.
However, the differences are relatively small; thus, the performance of predictive and periodic triggering is almost equal in the experiments.
In this case, the authors allowed for two communication slots per round for predictive and three for periodic triggering.
In a second experiment, one of the simulated systems was fixed at a specific position, \ie the part of \(v_i\) corresponding to the cart position had a non-zero mean, while the estimators still used~\eqref{eqn:state_est}.
Then, this system required more communication slots, and hence, the performance of predictive triggering was significantly better than the performance of periodic triggering.

The general setting in \cite{mager2021scaling} is slightly different since it is not a pure estimation scenario.
This indicates that the results obtained herein are applicable on a more general level.

The qualitative outcome of this analysis, that event-based mechanisms are most effective when we have disturbances or heterogeneous systems, is intuitive.
Further, Proposition~\ref{prop:per} and Lemma~\ref{lem:err_pred} provide a way to quantify it, which can be used to switch at runtime between event-based and periodic communication.

\section{Empirical evaluation}
\label{sec:eval}

We demonstrate the differences between predictive triggering and periodic round-robin communication in two examples.
First, we consider a synthetical example as a proof of concept. 
Then we turn to a setting with simulated cart-pole systems.

\subsection{Synthetical example}

For the synthetical example, we consider 20 homogeneous, linear, time-invariant systems as in~\eqref{eqn:sys}.
We sample the entries of \(A_i\) uniformly from the interval \([-1, 1]\), set \(\Sigma_i=\sigma_iI_4\), with \(\sigma_i=1\) for all \(i\) and \(I_4\) the \(4\times4\) identity matrix, \(K_\mathrm{per}=5\), \(K_\mathrm{pred}=2\), and \(\delta=0.1\).
After 100 iterations, we set \(\sigma_\mathrm{N}=0.5\) and after another 100 iterations, we set \(\sigma_i=0.2\) for all \(i\ge5\).

We compare the performance of periodic round-robin communication, predictive triggering, and an adaptive scheme.
We assume that we can detect the change in the dynamics and re-identify the system using, for instance, the algorithm proposed by \cite{solowjow2020event}.
This allows all estimators to use the actual system matrices for their state estimates.
Whenever the dynamics change, we also compare the errors in Proposition~\ref{prop:per} and Lemma~\ref{lem:err_pred}.
In the beginning, the systems are homogeneous.
This makes it beneficial to use the higher number of slots of periodic communication. 
Thus, the adaptive scheme equals the periodic one for the first 100 iterations.
After \(\sigma_\mathrm{N}\) changes, it switches to predictive triggering and allocates more slots to system \(N\).
After the second change, the adaptive scheme reverts to periodic communication.

\begin{figure}
\centering
\input{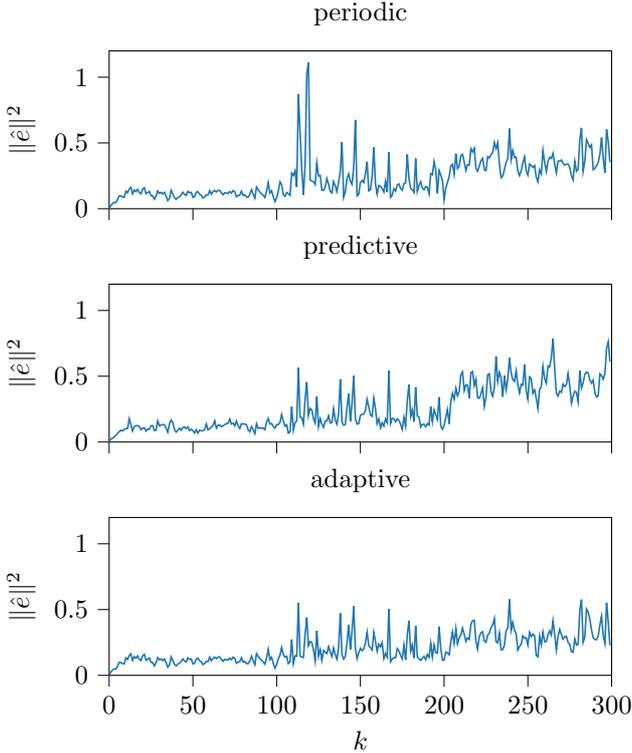}
\caption{Mean of the quadratic error of 20 systems with the different triggering schemes. \capt{In the beginning, round-robin communication and predictive triggering have similar performance. After the change in \(\sigma_\mathrm{N}\), predictive triggering can better serve the need of the heterogeneous, while when 15 systems simultaneously change \(\sigma_i\), the higher amount of communication slots of round-robin communication is more beneficial. Adaptive triggering always chooses the superior scheme.}}
\label{fig:full_traj}
\end{figure}

In \figref{fig:full_traj}, we show the mean of the quadratic error for each time step.
We see that in the beginning, when the process noise is relatively small, both periodic and predictive triggering have similar errors.
This changes after 100 iterations, when we set \(\sigma_\mathrm{n}=0.5\).
Now we have one system with a significantly greater demand for communication.
Predictive triggering can account for that, leading to smaller errors.
After setting \(\sigma_i=0.2\) for all \(i\ge5\), this changes.
Now there are more systems with a higher communication demand than predictive triggering can serve.
Thus, round-robin communication becomes the preferred option and can keep the errors lower.
The adaptive scheme always chooses the superior option.

\subsection{Cart-poles}

Next, we consider the widely studied cart-pole system.
We simulate 20 cart-poles using the Mujoco \citep{todorov2012mujoco} physics engine.
We assume that in the beginning, all systems have identical physical parameters but divide the mass of five carts by three after 100 time steps.
The systems are controlled by a manually tuned state feedback controller, and we identify system matrices via least-squares both before and after the change in the dynamics.
To make the task for the remote state estimators more challenging, we add a sine wave to the input of the cart-poles.
In this example, we set \(K_\mathrm{per}=5\), \(K_\mathrm{pred}=4\), and \(\delta=0.01\).
In the beginning, again, periodic triggering has the lower worst-case error, while after the dynamics change, it is predictive triggering.

\begin{figure}
\centering
\input{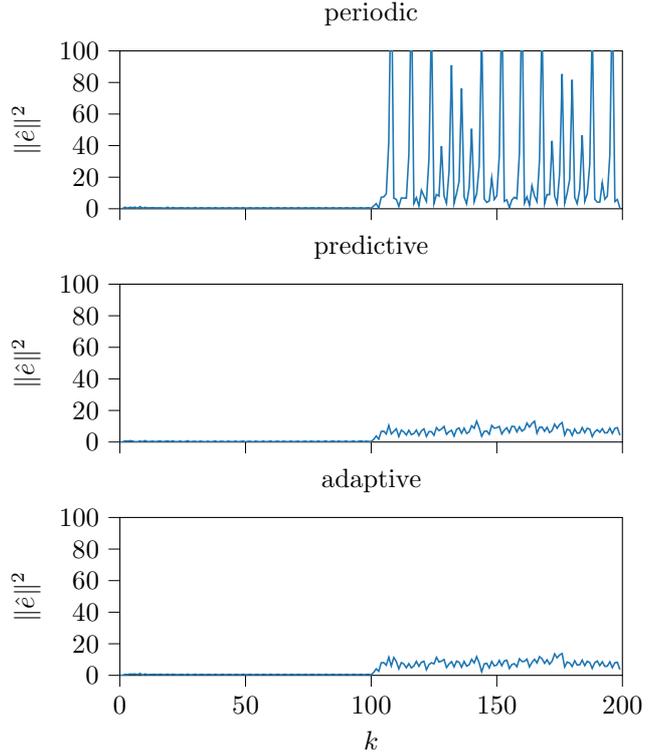}
\caption{Mean of the quadratic error of 20 cart-pole systems with the different triggering schemes. \capt{In the beginning, the errors of all schemes are almost identical. After changing the dynamics of five systems, predictive triggering outperforms periodic triggering. Again, the adaptive scheme always chooses the superior option.}}
\label{fig:cart-pole}
\end{figure}

We show the mean of the quadratic error for the three schemes in \figref{fig:cart-pole}.
When the multi-agent system is homogeneous, we can see that the difference between the different triggering schemes is negligible.
However, after the change in the dynamics, predictive triggering significantly outperforms periodic triggering.
Thus, also here it makes sense to evaluate which triggering scheme to choose online in case the system dynamics change.

\section{Discussion}

This paper investigated the trade-off of making informed communication decisions through event-based methods at the expense of using part of the bandwidth for scheduling purposes versus uninformed periodic schedules that can use the total bandwidth to transmit control messages.
We derived a criterion that can be used to check which sampling scheme is superior for a given multi-agent system and communication network.

The analysis in this paper is carried out for a specific event-based method and defines a criterion based on a worst-case error bound.
This is only a starting point for further research in this area.
Instead of investigating the worst-case performance, one could also study an average performance measure, for instance.
Another viable criterion could be robustness indices.
Event-based methods allow for online scheduling and, therefore, can immediately react to disturbances.
However, when a large number of systems is affected by a disturbance, \eg a wind gust hitting a swarm of drones, periodic triggering may still be superior due to the higher number of messages per round.

This paper investigates a setting where we schedule available slots to the agents to avoid overloading the bandwidth.
An orthogonal approach would be to let agents transmit information using an event-based mechanism and accept the risk of messages not being delivered if too many systems try to communicate simultaneously. 
This is often called decentralized scheduling and has, for instance, been studied by \cite{leong2016sensor}.
Instead of different number of slots per round, we would then have different message loss probabilities for event-based and periodic communication.
These could be traded off in a similar way as we have done in this paper.

In this work, we have studied the performance of a state-of-the-art event-based communication method and compared it to a relatively simple periodic triggering scheme.
In particular, the periodic scheme always allocates the same amount of bandwidth to each system. 
An alternative is to devise schemes where the amount of slots a system receives depends on its dynamics.
Then, the scheme could be adapted if the dynamics change, as in our evaluation example.
Developing such schemes is left for future work.

% \begin{ack}
% Place acknowledgments here.
% \end{ack}

\bibliography{ifacconf}            

\begin{thebibliography}{23}
\providecommand{\natexlab}[1]{#1}
\providecommand{\url}[1]{\texttt{#1}}
\providecommand{\urlprefix}{URL }
\expandafter\ifx\csname urlstyle\endcsname\relax
  \providecommand{\doi}[1]{doi:\discretionary{}{}{}#1}\else
  \providecommand{\doi}{doi:\discretionary{}{}{}\begingroup
  \urlstyle{rm}\Url}\fi

\bibitem[{Ara{\'u}jo et~al.(2013)Ara{\'u}jo, Mazo, Anta, Tabuada, and
  Johansson}]{araujo2013system}
Ara{\'u}jo, J., Mazo, M., Anta, A., Tabuada, P., and Johansson, K.H. (2013).
\newblock System architectures, protocols and algorithms for aperiodic wireless
  control systems.
\newblock \emph{IEEE Transactions on Industrial Informatics}, 10(1), 175--184.

\bibitem[{Balaghi et~al.(2018)Balaghi, Antunes, Mamduhi, Hirche
  et~al.}]{balaghi2018decentralized}
Balaghi, M.H., Antunes, D.J., Mamduhi, M.H., Hirche, S., et~al. (2018).
\newblock A decentralized consistent policy for event-triggered control over a
  shared contention-based network.
\newblock In \emph{IEEE Conference on Decision and Control}.

\bibitem[{Baumann et~al.(2019{\natexlab{a}})Baumann, Mager, Zimmerling, and
  Trimpe}]{baumann2019control}
Baumann, D., Mager, F., Zimmerling, M., and Trimpe, S. (2019{\natexlab{a}}).
\newblock Control-guided communication: Efficient resource arbitration and
  allocation in multi-hop wireless control systems.
\newblock \emph{IEEE Control Systems Letters}, 4(1), 127--132.

\bibitem[{Baumann et~al.(2019{\natexlab{b}})Baumann, Solowjow, Johansson, and
  Trimpe}]{baumann2019event}
Baumann, D., Solowjow, F., Johansson, K.H., and Trimpe, S.
  (2019{\natexlab{b}}).
\newblock Event-triggered pulse control with model learning (if necessary).
\newblock In \emph{American Control Conference}, 792--797.

\bibitem[{Dolk et~al.(2017)Dolk, Ploeg, and Heemels}]{dolk2017event}
Dolk, V.S., Ploeg, J., and Heemels, W.M.H. (2017).
\newblock Event-triggered control for string-stable vehicle platooning.
\newblock \emph{IEEE Transactions on Intelligent Transportation Systems},
  18(12).

\bibitem[{Gr{\"u}ne et~al.(2014)Gr{\"u}ne, Hirche, Junge, Koltai, Lehmann,
  Lunze, Molin, Sailer, Sigurani, St{\"o}cker, and Wirth}]{grune14event}
Gr{\"u}ne, L., Hirche, S., Junge, O., Koltai, P., Lehmann, D., Lunze, J.,
  Molin, A., Sailer, R., Sigurani, M., St{\"o}cker, C., and Wirth, F. (2014).
\newblock Event-based control.
\newblock In J.~Lunze (ed.), \emph{Control Theory of Digitally Networked
  Dynamic Systems}, 169--261. Springer.

\bibitem[{Heemels et~al.(2012)Heemels, Johansson, and
  Tabuada}]{heemels2012introduction}
Heemels, W.P.M.H., Johansson, K.H., and Tabuada, P. (2012).
\newblock An introduction to event-triggered and self-triggered control.
\newblock In \emph{IEEE Conference on Decision and Control}, 3270--3285.

\bibitem[{Hespanha et~al.(2007)Hespanha, Naghshtabrizi, and
  Xu}]{hespanha2007survey}
Hespanha, J.P., Naghshtabrizi, P., and Xu, Y. (2007).
\newblock A survey of recent results in networked control systems.
\newblock \emph{Proceedings of the IEEE}, 95(1), 138--162.

\bibitem[{Lemmon(2010)}]{lemmon11networked}
Lemmon, M. (2010).
\newblock Event-triggered feedback in control, estimation, and optimization.
\newblock In A.~Bemporad, M.~Heemels, and M.~Johansson (eds.), \emph{Networked
  Control Systems}, volume 406 of \emph{Lecture Notes in Control and
  Information Sciences}, 293--358. Springer.

\bibitem[{Leong et~al.(2016)Leong, Dey, and Quevedo}]{leong2016sensor}
Leong, A.S., Dey, S., and Quevedo, D.E. (2016).
\newblock Sensor scheduling in variance based event triggered estimation with
  packet drops.
\newblock \emph{IEEE Transactions on Automatic Control}, 62(4), 1880--1895.

\bibitem[{Ma and Lu(2018)}]{ma2018efficient}
Ma, Y. and Lu, C. (2018).
\newblock Efficient holistic control over industrial wireless sensor-actuator
  networks.
\newblock In \emph{IEEE International Conference on Industrial Internet},
  89--98.

\bibitem[{Mager et~al.(2021)Mager, Baumann, Herrmann, Trimpe, and
  Zimmerling}]{mager2021scaling}
Mager, F., Baumann, D., Herrmann, C., Trimpe, S., and Zimmerling, M. (2021).
\newblock Scaling beyond bandwidth limitations: Wireless control with stability
  guarantees under overload.
\newblock \emph{ACM Transactions on Cyber-Physical Systems}, 6(3), 30.

\bibitem[{Mamduhi et~al.(2017)Mamduhi, Molin, Toli{\'c}, and
  Hirche}]{mamduhi2017error}
Mamduhi, M.H., Molin, A., Toli{\'c}, D., and Hirche, S. (2017).
\newblock Error-dependent data scheduling in resource-aware multi-loop
  networked control systems.
\newblock \emph{Automatica}, 81, 209--216.

\bibitem[{Mastrangelo et~al.(2019)Mastrangelo, Baumann, and
  Trimpe}]{mastrangelo2019predictive}
Mastrangelo, J.M., Baumann, D., and Trimpe, S. (2019).
\newblock Predictive triggering for distributed control of resource constrained
  multi-agent systems.
\newblock \emph{IFAC Workshop on Distributed Estimation and Control in
  Networked Systems}, 52(20), 79--84.

\bibitem[{Mazo~Jr et~al.(2010)Mazo~Jr, Anta, and Tabuada}]{mazo2010iss}
Mazo~Jr, M., Anta, A., and Tabuada, P. (2010).
\newblock An {ISS} self-triggered implementation of linear controllers.
\newblock \emph{Automatica}, 46(8), 1310--1314.

\bibitem[{Miskowicz(2016)}]{miskowicz15event}
Miskowicz, M. (2016).
\newblock \emph{Event-Based Control and Signal Processing}.
\newblock CRC Press.

\bibitem[{Molin and Hirche(2011)}]{molin2011optimal}
Molin, A. and Hirche, S. (2011).
\newblock Optimal design of decentralized event-triggered controllers for
  large-scale systems with contention-based communication.
\newblock In \emph{IEEE Conference on Decision and Control and European Control
  Conference}, 4710--4716.

\bibitem[{Ramesh et~al.(2016)Ramesh, Sandberg, and
  Johansson}]{ramesh2016performance}
Ramesh, C., Sandberg, H., and Johansson, K.H. (2016).
\newblock Performance analysis of a network of event-based systems.
\newblock \emph{IEEE Transactions on Automatic Control}, 61(11), 3568--3573.

\bibitem[{Solowjow and Trimpe(2020)}]{solowjow2020event}
Solowjow, F. and Trimpe, S. (2020).
\newblock Event-triggered learning.
\newblock \emph{Automatica}, 117, 109009.

\bibitem[{Todorov et~al.(2012)Todorov, Erez, and Tassa}]{todorov2012mujoco}
Todorov, E., Erez, T., and Tassa, Y. (2012).
\newblock Mujoco: A physics engine for model-based control.
\newblock In \emph{IEEE/RSJ International Conference on Intelligent Robots and
  Systems}, 5026--5033.

\bibitem[{Trimpe and D'Andrea(2011)}]{trimpe2011experimental}
Trimpe, S. and D'Andrea, R. (2011).
\newblock An experimental demonstration of a distributed and event-based state
  estimation algorithm.
\newblock In \emph{IFAC World Congress}, 8811--8818.

\bibitem[{Velasco et~al.(2003)Velasco, Fuertes, and Marti}]{velasco2003self}
Velasco, M., Fuertes, J., and Marti, P. (2003).
\newblock The self triggered task model for real-time control systems.
\newblock In \emph{Work-in-Progress Session of the IEEE Real-Time Systems
  Symposium}, 67--70.

\bibitem[{Wang and Lemmon(2009)}]{wang2009self}
Wang, X. and Lemmon, M.D. (2009).
\newblock Self-triggered feedback control systems with finite-gain
  $\mathcal{L}_{2}$ stability.
\newblock \emph{IEEE Transactions on Automatic Control}, 54(3), 452--467.

\end{thebibliography}
\end{document}